\documentclass[11pt]{article}

\usepackage[utf8]{inputenc}
\usepackage[T1]{fontenc}
\usepackage{amsmath,amssymb,amsthm}
\usepackage{graphicx}
\usepackage{hyperref}
\usepackage{geometry}
\hypersetup{
    colorlinks=true,
    linkcolor=blue,
    citecolor=red,
    urlcolor=blue
}

\title{Research of the Behavior of the Effective Potential in Systems with Phase Transitions through the Prism of A--D--E Type Singularities}

\author{T. V. Obikhod\\
\small Institute for Nuclear Research, National Academy of Sciences of Ukraine,\\
\small pr. Nauky 47, 03680 Kyiv, Ukraine\\
\small \texttt{obikhod@kinr.kiev.ua}
}

\begin{document}

\maketitle

\begin{abstract}
Detecting a scalar singlet interacting through the Higgs portal demands a pivot from conventional particle detection strategies to a comprehensive examination of the effective potential's landscape. The presence, intensity, and first-order nature of the electroweak phase transition are dictated by the critical manifold, with its universal traits encapsulated in the Milnor number $\mu$---the dimensionality of the local Jacobian algebra. Throughout the parameter space consistent with experimental observations, the portal potential exhibits a non-simple singularity with $\mu = 9$, maintaining topological stability amid substantial fluctuations in mixing angle, singlet mass, and cubic interactions. High-precision assessments of the Higgs trilinear self-coupling ($\kappa_\lambda$), the uniform rescaling of Higgs couplings ($c_H$), and the stochastic gravitational-wave spectrum ($\Omega_{\mathrm{GW}}$) collectively delineate the catastrophe, extending beyond mere mass matrix analysis. Projections for 2027--2040 collider and LISA capabilities indicate that no viable region supporting a strong first-order transition will evade scrutiny; thus, the singlet will either be identified or conclusively dismissed via direct interrogation of the electroweak vacuum's critical structure.
\end{abstract}
\vspace{0.5cm}

\section{Introduction}
\label{sec:intro}

The classification of isolated singularities and their relation to symmetry has played a central role in the development of modern theoretical physics. Among the most remarkable structures emerging in this context is the ADE classification, which unifies seemingly distinct mathematical objects such as simple Lie algebras, Dynkin diagrams, and simple hypersurface singularities. This unification has profound implications for quantum field theory, string theory, and models of physics beyond the Standard Model (BSM).

In many extensions of the Standard Model, the scalar sector is enlarged to include additional Higgs fields, moduli, or effective degrees of freedom arising from ultraviolet completions. The vacuum structure of such theories is determined by the critical points of scalar potentials, whose local behaviour can often be described in terms of isolated singularities. When these singularities are simple, their local form is completely classified by the ADE series \cite{1.}. As a consequence, ADE-type potentials provide a natural organizing principle for understanding universality and symmetry enhancement in BSM scalar dynamics.

A key invariant characterizing an isolated singularity is the Milnor number, which measures the dimension of the local deformation space of the potential. From a physical perspective, the Milnor number counts the number of independent relevant deformations, determines the structure of the vacuum moduli space, and controls the spectrum of light degrees of freedom near critical points. In supersymmetric settings, it is directly related to the number of chiral primary operators in the associated Landau--Ginzburg or superconformal field theory, while in geometric constructions it corresponds to the number of vanishing cycles arising in the resolution of singularities.

The appearance of ADE structures is not limited to abstract classification results. In string theory, ADE singularities of compactification manifolds give rise to non-Abelian gauge symmetries and exceptional groups, providing a geometric origin for gauge sectors beyond the Standard Model. In effective field theory, ADE-type scalar potentials describe universal critical behaviour near phase transitions and symmetry-breaking points, linking singularity theory to early-Universe cosmology and vacuum dynamics. These connections suggest that ADE classification may serve as a powerful constraint on viable BSM scenarios, particularly in theories with strongly constrained scalar sectors.

The purpose of this paper is to investigate the role of ADE classification in the study of scalar potentials relevant for physics beyond the Standard Model, with particular emphasis on the interpretation of the Milnor number as a physically meaningful quantity. We analyze how ADE-type singularities organize the space of effective potentials, constrain their deformation patterns, and provide a bridge between algebraic, geometric, and field-theoretic descriptions of BSM physics. Our results highlight the usefulness of singularity theory as a systematic framework for exploring vacuum structure and universality in extensions of the Standard Model.

\section{Conceptual fragility of the Standard-Model Higgs potential}
\label{sec:fragility}

In 1964, three groups of theorists independently resolved a key obstacle in unifying electromagnetism and the weak force: the apparent incompatibility between gauge invariance—which requires massless force carriers—and the need for massive mediators to explain the weak interaction’s short range. Their solution, now known as the Brout--Englert--Higgs (BEH) mechanism, showed how spontaneous symmetry breaking could give mass to vector bosons while preserving renormalisability. Peter Higgs highlighted the existence of a massive scalar particle—the Higgs boson—while Guralnik, Hagen, and Kibble provided a general proof of the mechanism’s consistency with quantum field theory principles. Steven Weinberg and Abdus Salam (building on Glashow’s work) incorporated the BEH mechanism into the electroweak $\mathrm{SU}(2)_L \times \mathrm{U}(1)_Y$ theory, predicting the $W$ and $Z$ boson mass ratio later confirmed experimentally. The Higgs boson itself was finally discovered on 4~July~2012 at the LHC, with a mass of $125.38\,\mathrm{GeV}$—within the narrow range allowing vacuum stability up to the Planck scale, a feature that still shapes theories beyond the Standard Model.

The BEH mechanism, however, relies on an unexplained assumption: the Higgs potential’s quadratic term must have a negative coefficient ($\mu^2 < 0$). No internal symmetry dictates this sign, nor does the theory predict the magnitude of $\mu^2$—instead, we infer it from the measured vacuum expectation value $v \approx 246\,\mathrm{GeV}$, treating both $v$ and the Higgs self-coupling $\lambda$ as empirical inputs. This conceptual fragility echoes the 1950 Ginzburg--Landau theory of superconductivity, which similarly postulated a phenomenological potential without deriving it from first principles. In 1950, Ginzburg and Landau wrote the free-energy density of a superconductor as
\[
F = \alpha |\Psi|^2 + \frac{1}{2}\beta |\Psi|^4 + \frac{1}{2m^*} |(\nabla - 2ie\mathbf{A})\Psi|^2 + \frac{B^2}{2\mu_0},
\]
with $\alpha \propto (T - T_c)$. The sign of $\alpha$ is not derived; it is imported from experiment. When $T < T_c$, the coefficient turns negative, the $\mathrm{U}(1)$ gauge symmetry of electromagnetism is spontaneously broken inside the material, and the photon acquires an effective mass—manifested as London’s penetration depth. The formal parallel to the Higgs potential
\[
V(H) = \mu^2 |H|^2 + \lambda |H|^4, \qquad \mu^2 < 0,
\]
is unmistakable. Both are Landau expansions with order parameters, symmetry breaking, and energy gaps—yet neither explains the origin of the quadratic sign flip.

Superconductivity’s resolution came via Bardeen--Cooper--Schrieffer (BCS): the negativity of $\alpha$ emerges dynamically from attractive, phonon-mediated electron pairing. Below the Debye frequency $\Omega_D$, a four-fermion coupling $G > 0$ becomes relevant under renormalization-group flow, instigating a Cooper instability at $\Delta \simeq 2\Omega_D \exp[-1/(N(0)G)]$. Integrating out high-energy electrons yields the Ginzburg--Landau coefficients:
\[
\alpha = N(0)\frac{T - T_c}{T_c}, \qquad \beta = \frac{7\zeta(3)N(0)}{8\pi^2 T_c^2},
\]
where the sign shift of $\alpha$ is dictated by the microscopic coupling $G$ and the density of states $N(0)$ at the Fermi surface.

In contrast, the origin of electroweak symmetry breaking (EWSB) and the sign of $\mu^2$ remain enigmatic. While the Standard Model (SM) assumes a negative mass-squared, theories beyond the SM (BSM) offer dynamical mechanisms:
\begin{itemize}
    \item \textbf{Technicolor} composites the Higgs via a new strong force; EWSB arises from a condensate $\langle \bar{Q} Q \rangle \neq 0$, analogous to chiral symmetry breaking in QCD, with the Higgs as a pseudo-Nambu--Goldstone boson and its potential generated by strong dynamics.
    \item \textbf{Supersymmetry}, particularly the MSSM, radiatively triggers EWSB: the large top Yukawa coupling drives the up-type Higgs mass squared negative via renormalization from high scales, naturally yielding $\mu^2 < 0$ and electroweak breaking.
    \item \textbf{Extra-dimensional models} (e.g., Randall--Sundrum or Gauge-Higgs Unification) recast the Higgs as the fifth component of a gauge field; the tree-level potential is forbidden by bulk gauge symmetry and is instead generated radiatively via Casimir energy and loop effects, with the vacuum structure shaped by geometry and boundary conditions.
\end{itemize}

\begin{figure}[t]
\centering
\includegraphics[width=0.6\textwidth]{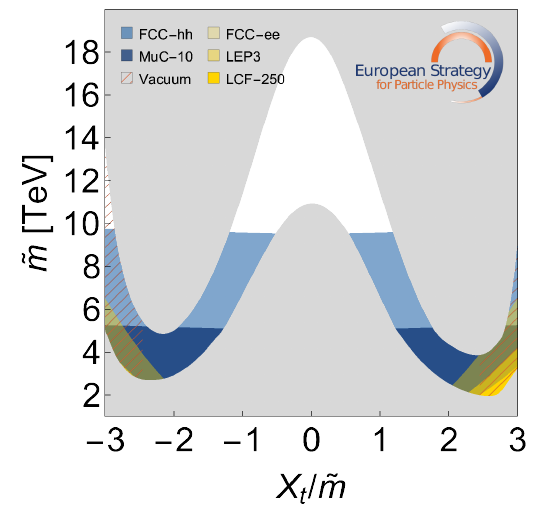}
\caption{Parameter space for $m_h = 125\,\mathrm{GeV}$ \cite{2.}. The abscissa is the mixing parameter $X_t \equiv A_t - \mu \cot\beta$, the ordinate is the average stop soft mass $\tilde{m} \equiv \sqrt{m_{Q_3} m_{U_3}}$. The grey band marks the region where the two-loop NNLL calculation of \texttt{FeynHiggs 2.19.0} \cite{3.} reproduces $m_h = 125\,\mathrm{GeV}$ within $2\sigma$. The parabolic shape follows $\Delta m_h^2 \simeq \frac{3 y_t^2}{8\pi^2} \tilde{m}^2 \left(1 - \frac{X_t^2}{6\tilde{m}^2}\right)^2 \ln(\tilde{m}^2/m_t^2)$.}
\label{fig:stop}
\end{figure}

Over the next two decades, precision measurements of Higgs couplings, longitudinal $W$ scattering, and triple Higgs production will probe the Higgs potential at the 5--10\% level. Any deviation from the SM prediction $\lambda \approx 0.129$ would indicate the presence of new heavy states contributing to the Higgs self-energy—akin to phonon-mediated interactions in BCS theory—while agreement would push the scale of new physics higher, exacerbating the naturalness problem. The core goal is to determine the sign and magnitude of the Higgs mass parameter $\mu^2$ in the presence of possible beyond-SM effects. A key example is the supersymmetric stop sector, where top/stop loops simultaneously shape the Higgs mass and electroweak precision observables. Future colliders will thus test three interlinked aspects of the same vacuum instability, as illustrated in Fig.~\ref{fig:stop}.

Thus, the $\{\tilde{m}, X_t\}$ plane maps ``vacuum solutions'' to $\mu^2 = -\lambda v^2$: natural in one region ($\tilde{m} \sim 1\,\mathrm{TeV}$, large mixing), fine-tuned in another (multi-TeV stops), and unstable in a third. Measuring the boundary between these regions discerns the microscopic theory responsible for the sign of $\mu^2$. An extra scalar can resolve this ambiguity. Adding a real singlet $S$ via
\[
\Delta \mathcal{L} = \kappa S |H|^2 + \frac{1}{2}(\partial S)^2 - \frac{1}{2} m_S^2 S^2 - \lambda_S S^4,
\]
introduces a new axis: the sign and size of $\mu^2 = -\lambda v^2 + \kappa \langle S \rangle$ now depend on the vacuum expectation value of an additional scalar.

Scalar singlets—particularly Standard Model gauge singlets—are weakly constrained because they couple feebly and leave only subtle traces in collider data. However, they can modify the Higgs effective potential through loop corrections, even if too heavy or weakly coupled to be produced directly. Upcoming Higgs factories will probe these tiny distortions with unprecedented precision, offering a unique window into minimal extensions of the SM. To illustrate this potential, consider the simplest benchmark: a real singlet $S$ coupled only via the Higgs portal,
\begin{equation}
V(h, s) = -\frac{\mu_H^2}{2} h^2 + \frac{\lambda}{4} h^4 
          -\frac{\mu_s^2}{2} s^2 + \frac{b_3}{3} s^3 + \frac{b_4}{4} s^4 
          + \frac{a_1}{4} h^2 s + \frac{a_2}{4} h^2 s^2 + b_1s\, .
\end{equation}
where
\begin{itemize}
    \item $h$ is the real scalar field (the physical Higgs component),
    \item $S$ is a real singlet scalar,
    \item $\mu_H^2, \mu_S^2$ are coefficients of the quadratic terms,
    \item $\lambda$ is the Higgs quartic coupling,
    \item $a_1$ is a linear-in-$S$ mixing term with $h^2$,
    \item $a_2$ is the portal quartic coupling ($h^2 S^2$),
    \item $b_3, b_4, b_1$ are self-interaction terms of $S$.
\end{itemize}

After electroweak symmetry breaking, the singlet $S$ can acquire a vacuum expectation value $w$, leading to tree-level mixing with the SM Higgs $h$. This modifies Higgs couplings and enables resonant production of a new scalar state decaying into SM particles. If a $\mathbb{Z}_2$ symmetry ($S \to -S$) is imposed, odd terms vanish ($a_1 = b_3 = 0$), forcing $w = 0$. The singlet then remains inert at tree level but still affects Higgs physics through loop corrections. The physical singlet mass is $m_s^2 = -\mu_s^2 + \frac{\lambda_{hs}}{2} v^2$, and for $m_s < m_h/2$, the decay $h \to ss$ occurs with coupling $g_{hss} = \lambda_{hs} v$, yielding a width proportional to $(\lambda_{hs} v)^2$. Crucially, even under an exact $\mathbb{Z}_2$, the singlet induces measurable loop-level shifts in Higgs couplings. Thus, a scalar singlet is the simplest SM extension that can leave observable signatures—either via tree-level mixing or through quantum effects—making it a prime target for precision Higgs studies at future colliders.

\begin{figure}[t]
\centering
\includegraphics[width=0.55\textwidth]{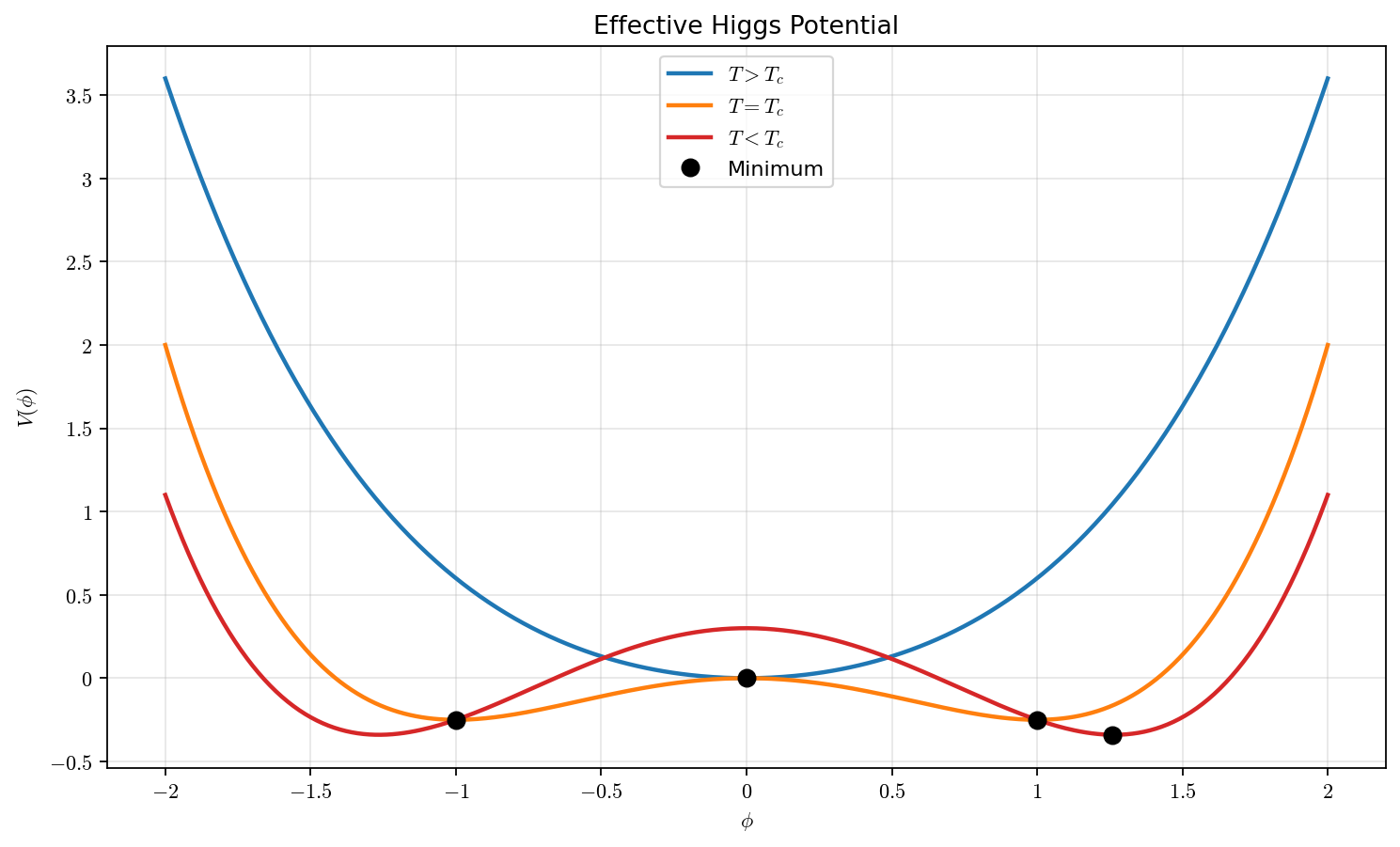}
\caption{The SM transition as a smooth crossover and FOPT that involves the coexistence of two phases.}
\label{fig:fopt}
\end{figure}

Lattice studies show that, for the observed Higgs mass $ m_h\simeq 125\,\mathrm{GeV}$, the SM electroweak phase transition (EWPT) is a smooth crossover rather than a true phase transition. In contrast, a first-order phase transition (FOPT) \cite{4.} is characterised by the coexistence of symmetric and broken phases separated by a potential barrier, leading to bubble nucleation and expansion (Fig.~\ref{fig:fopt}). FOPT is phenomenologically important since it enables:
\begin{itemize}
    \item Electroweak baryogenesis, satisfying the Sakharov conditions via sphaleron processes, CP violation, and departure from thermal equilibrium;
    \item Stochastic gravitational wave production from bubble collisions and plasma turbulence;
    \item Signals of physics beyond the SM.
\end{itemize}
Since the SM predicts a crossover, researchers investigate extensions such as Two-Higgs-Doublet Models (2HDMs) or scalar singlets. These models modify the Higgs potential—often through large two-loop effects or additional cubic terms—to facilitate a strong FOPT. Extensions of the scalar sector can thus induce a strong first-order EWPT while predicting observable deviations in Higgs couplings, testable at future colliders.

\section*{3. Summary of Parametric Space}

\begin{figure}[ht]
\centering
\includegraphics[width=0.6\textwidth]{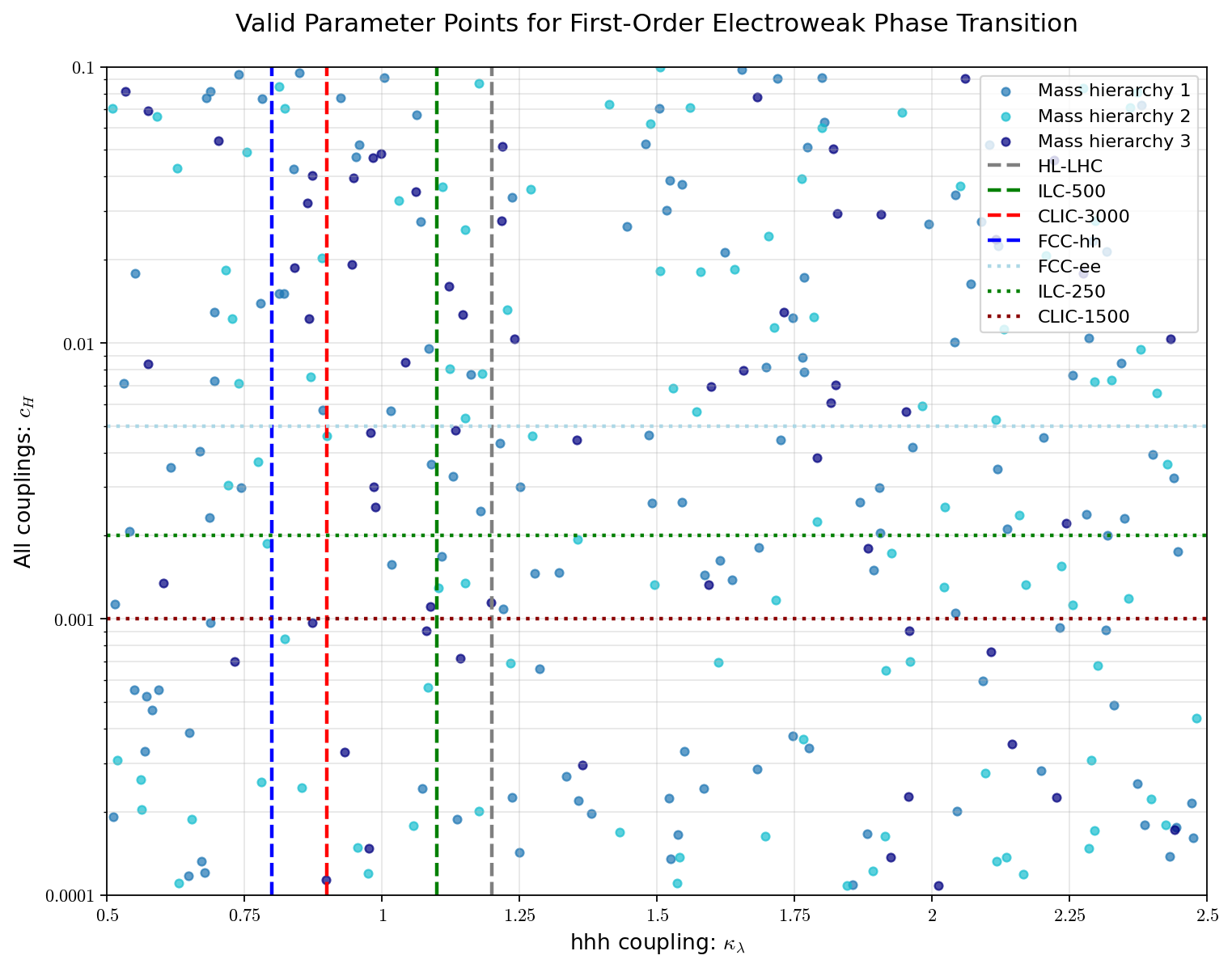}
\caption{Parameter space of the singlet scalar as a function of the universal shift to all Higgs couplings ($c_H$) and Higgs self-coupling ($\kappa_\lambda$).}
\label{fig:fig3}
\end{figure}

The SM Higgs potential changes smoothly from the symmetric to the broken phase (crossover). Adding a real singlet $S$ with renormalizable interactions
\[
V \supset \frac{1}{2} a_1 S |H|^2 + \frac{1}{2} a_2 S^2 |H|^2 + \frac{1}{3} b_3 S^3 + \frac{1}{4} b_4 S^4
\]
can build a potential barrier between the two vacua once the temperature drops below $T_c$. The barrier persists if at $T_c$ the order-parameter jump satisfies $v_c / T_c \geq 1$, the field-theory benchmark for a strong first-order phase transition (SFOPT) and successful electroweak baryogenesis (EWBG). Bubble nucleation during such a transition sources a gravitational wave (GW) spectrum peaking at milli-Hertz—precisely the LISA band. A sub-TeV gauge-singlet scalar converts EWSB to a FOPT. The same couplings imprint a universal shift on all Higgs couplings and a measurable trilinear deviation
\[
\kappa_\lambda = \frac{\lambda_{hhh}}{\lambda_{hhh}^{\text{SM}}}
\]
in the Higgs self-coupling. Figure~\ref{fig:fig3} shows that the parameter islands giving a strong FOPT cluster in the region $\kappa_\lambda \gtrsim 1.5$ and $c_H \gtrsim 0.1$—precisely the territory that future colliders plan to map.

The universal coupling shift ($c_H$) quantifies how the couplings of the observed Higgs boson ($h_1$) to all Standard Model (SM) particles are modified due to its mixing with a new scalar field (in this case, a singlet scalar $S$). The parameter $c_H$ is typically defined as a measure of this deviation, specifically related to the square of the sine of the mixing angle:
\[
c_H \approx \sin^2\theta.
\]
The mixing angle $\theta$ itself is a function of the potential parameters $(\mu^2, \lambda, a_1, a_2, b_3, b_4)$ and the vacuum expectation values (VEVs).

The density and distribution of these points reveal several physical insights:

\paragraph{Mass Hierarchies:} The legend distinguishes between different mass relations for the new scalar $m_{h_2}$:
\begin{itemize}
    \item Light blue: $m_{h_1} < m_{h_2} < 2m_{h_1}$
    \item Medium blue: $2m_{h_1} < m_{h_2} < 3m_{h_1}$ (likely indicating a transition region or specific resonance condition)
    \item Dark blue: $m_{h_2} > 2m_{h_1}$
\end{itemize}

\begin{figure}[ht]
\centering
\includegraphics[width=0.6\textwidth]{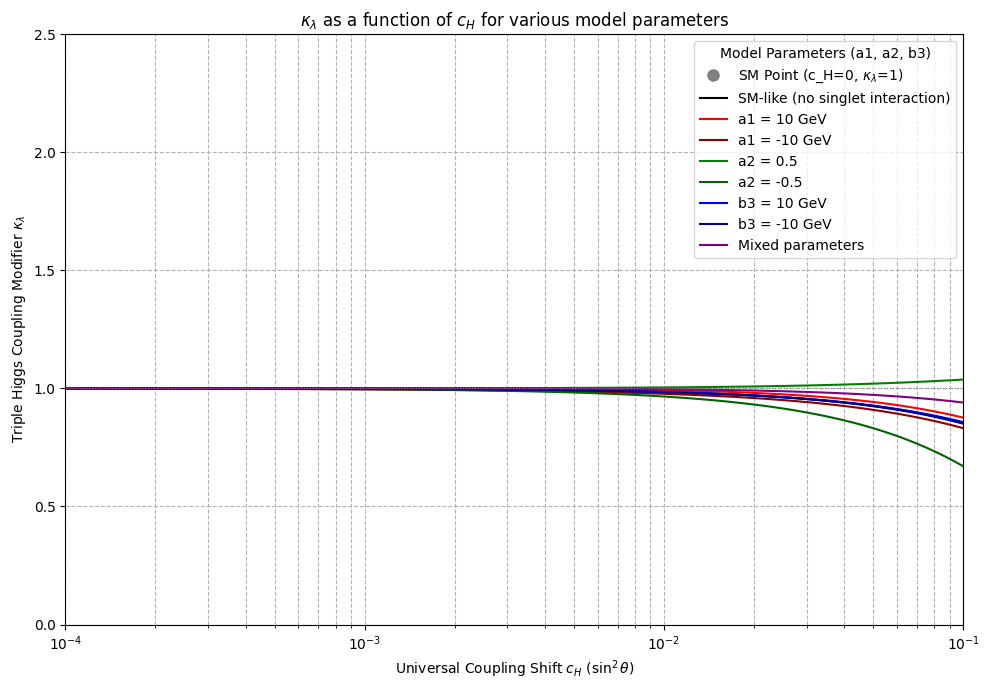}
\caption{$\kappa_\lambda$ versus $c_H$ for assorted singlet-model parameters.}
\label{fig:fig4}
\end{figure}

\paragraph{Correlation:} There is a clear ``V-shaped'' or ``funnel'' structure centered around $\kappa_\lambda \approx 1$. As the universal coupling shift $c_H$ decreases (moving down the y-axis), the required deviation in the self-coupling $\kappa_\lambda$ to maintain a first-order transition typically increases. The formula of their connection is the following
\[\kappa_{\lambda} = (1 - c_H)^{3/2} + \frac{1}{6\lambda_v} \left( \frac{3a_1}{2} (1 - c_H) \sqrt{c_H} + 3a_2 v c_H \sqrt{1 - c_H} + 2b_3 c_H^{3/2} \right) \ .\]
Figure~\ref{fig:fig4} shows how the Higgs self-coupling multiplier $\kappa_\lambda$ changes as the universal shift $c_H \approx \sin^2\theta$ is scanned, with each line corresponding to a different choice of $a_1$, $a_2$, and $b_3$.

\begin{figure}[ht]
\centering
\includegraphics[width=\textwidth]{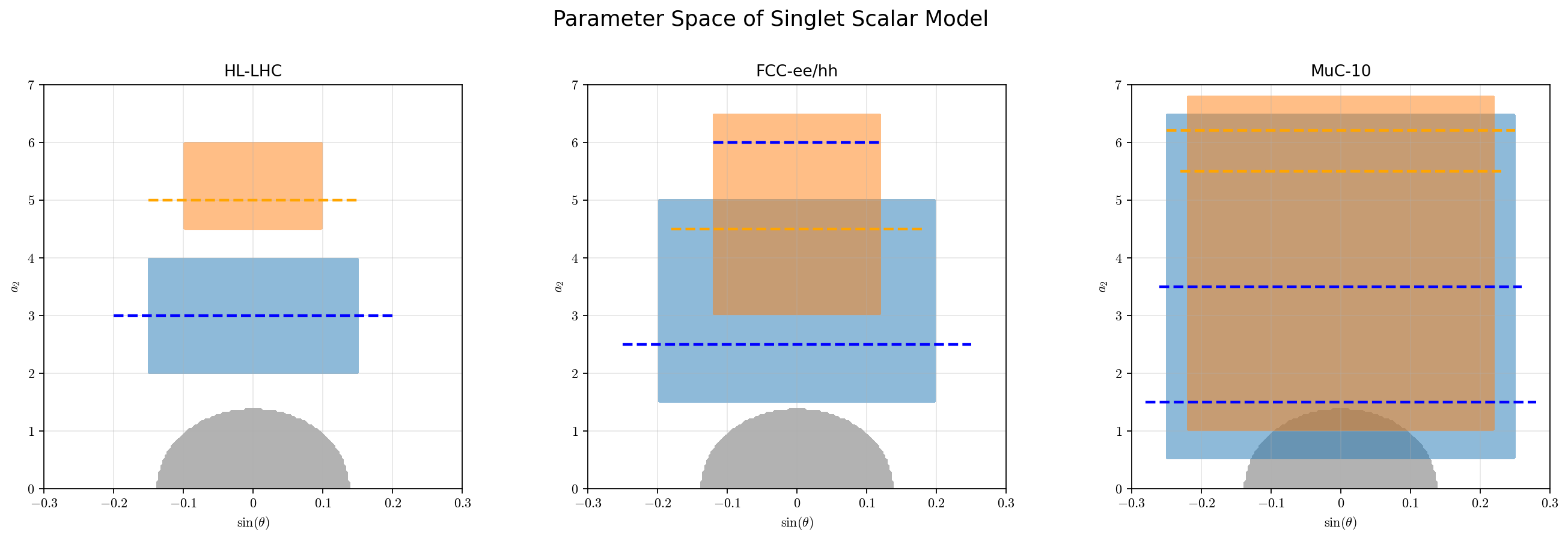}
\caption{Parameter space for the direct (solid regions) and indirect (dashed lines) constraints on the scalar field.}
\label{fig:fig5}
\end{figure}

Thus, the singlet scalar model provides a viable pathway for a first-order electroweak phase transition, but it requires measurable deviations in Higgs properties. Future colliders, such as FCC-ee and FCC-hh, are essential for probing the $c_H$ and $\kappa_\lambda$ values necessary to validate or exclude this mechanism for baryogenesis. The limits are categorized into direct and indirect searches (Figure~\ref{fig:fig5}).

\paragraph{Direct Searches (Solid Regions):}
\begin{itemize}
    \item $S \to ZZ$ (Blue): Highly sensitive to the mixing angle $\sin\theta$.
    \item $S \to hh$ (Orange): Searches for resonant production of Higgs boson pairs.
\end{itemize}

\paragraph{Indirect Constraints (Dashed Lines):}
\begin{itemize}
    \item Universal Coupling Shift ($c_H$): Any deviation from $\cos\theta = 1$ signals new physics. FCC-ee is particularly powerful here due to its high-precision Higgs factory capabilities.
    \item Higgs Self-coupling ($h^3$): Measurements of the trilinear Higgs coupling $\lambda_{hhh}$.
\end{itemize}

The grey shaded region represents the parameter space where a strong FOPT occurs.

\begin{figure}[ht]
\centering
\includegraphics[width=0.6\textwidth]{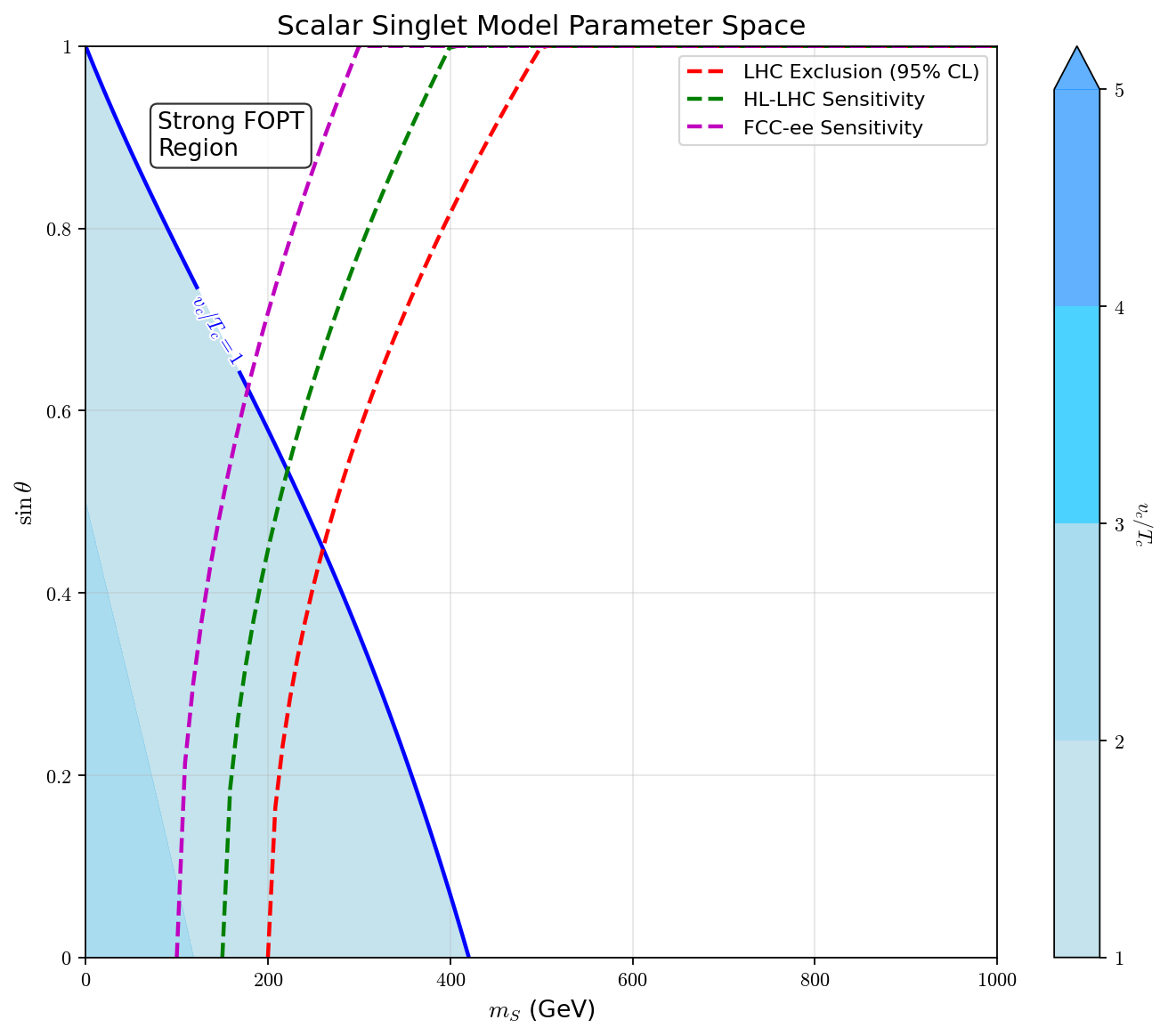}
\caption{Scalar-singlet parameter space in the $(m_S, \sin\theta)$ plane.}
\label{fig:fig6}
\end{figure}

Figure~\ref{fig:fig6} overlays the 95\% CL LHC exclusion, HL-LHC and FCC-ee sensitivities, and the band where the model drives a strong first-order electroweak phase transition.

The search for a scalar-singlet-driven FOPT relies on two complementary approaches:
\begin{itemize}
    \item \textbf{Direct Searches:} Production of the new scalar $S$ and its decay, e.g., $pp \to S \to ZZ$ or $pp \to S \to hh$, effective for $m_S > 2m_h$.
    \item \textbf{Indirect Searches:} Deviations in $\lambda_{hhh}$ and universal shifts from mixing $\sin\theta$.
\end{itemize}

\paragraph{The $Z_2$ Limit:} When $\sin\theta \to 0$, direct production of $S$ via mixing vanishes. In this regime, the transition must be probed via precision measurements of $\lambda_{hhh}$ and the $Zh$ production cross-section at lepton colliders.

The interplay between direct and indirect constraints is summarized as follows:
\begin{itemize}
    \item For $m_S \approx {600}{GeV}$, the region yielding a strong FOPT is increasingly squeezed by current LHC data but remains accessible to future high-precision experiments.
    \item The scalar singlet serves as a ``proxy'' for more complex extended sectors (like 2HDM). If a strong FOPT occurred in the early universe, the associated scalar must reside in a region that significantly modifies Higgs properties—making it a "no-lose" theorem for future collider precision.
\end{itemize}

The analysis concludes that a strong FOPT in the real scalar singlet model is highly correlated with detectable shifts in the Higgs sector, ensuring that future colliders will either discover the singlet or rule out its role in EWBG.

\begin{figure}[ht]
\centering
\includegraphics[width=\textwidth]{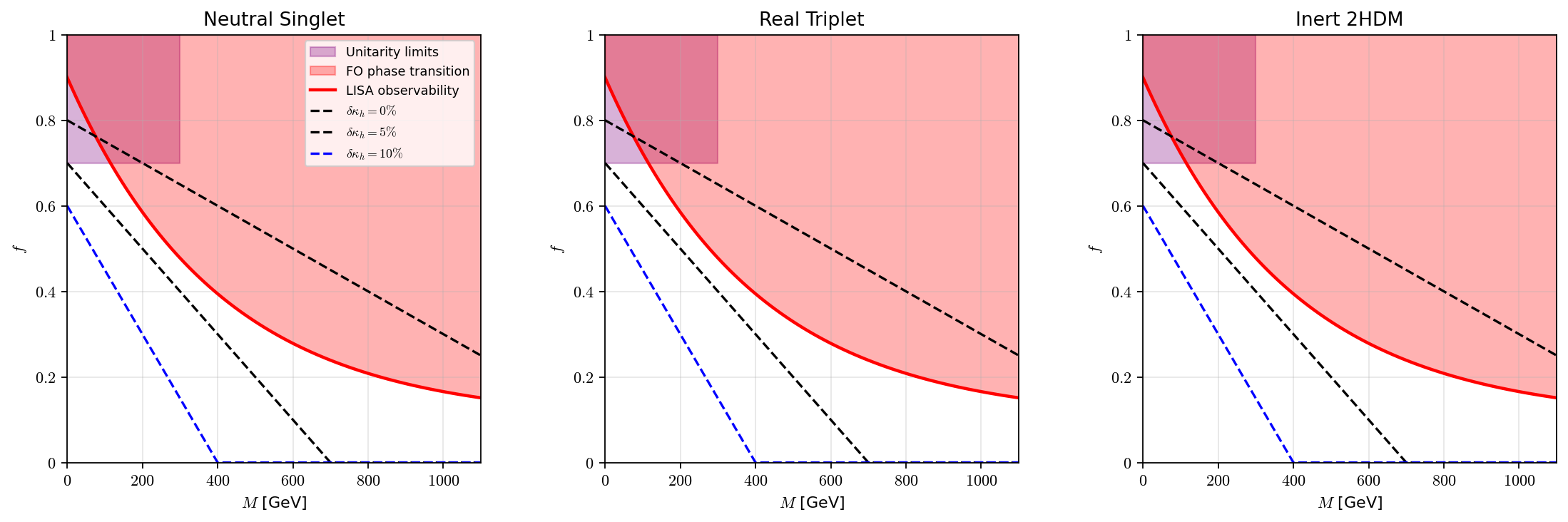}
\caption{Strong-first-order-phase-transition reach versus unitarity and LISA for $\mathbb{Z}_2$-symmetric scalars.}
\label{fig:fig7}
\end{figure}

Figure~\ref{fig:fig7} compares, for a neutral singlet, real triplet, and inert 2HDM models, the mass–mixing region producing a first-order transition (red), the part detectable by LISA (red line), and the 95\% CL exclusion that future Higgs-precision measurements will impose (dashed curves).

The parameter space is mapped using two primary variables:
\begin{itemize}
    \item \textbf{Physical Mass ($M$):} The pole mass of the new scalar particle, measured in GeV.
    \item \textbf{Mass Fraction ($f$):} The fraction of the scalar's mass squared that originates from EWSB. For a scalar $S$ with portal coupling $\lambda_{hS} |H|^2 |S|^2$, the physical mass is
    \[
    M^2 = \mu_S^2 + \lambda_{hS} v^2,
    \]
    and the fraction $f$ is defined as
    \[
    f = \frac{\lambda_{hS} v^2}{M^2}.
    \]
    A value $f = 1$ implies the mass is entirely from EWSB; $f \to 0$ means it is dominated by the bare mass $\mu_S$.
\end{itemize}

The plots in Fig.~\ref{fig:fig7} delineate critical boundaries:
\begin{itemize}
    \item \textbf{Perturbative Unitarity (Purple):} Where $\lambda_{hS}$ becomes non-perturbative.
    \item \textbf{Higgs Precision (Dashed):} New scalars modify Higgs couplings via loop effects.
    \item \textbf{EWPT (Red):} Region of strong FOPT.
    \item \textbf{LISA Sensitivity (Red Line):} Stochastic GW background from FOPT.
\end{itemize}

\paragraph{Comparison Across Models:}
\begin{itemize}
    \item \textit{Neutral Singlet:} Broad FOPT region requiring high $f$.
    \item \textit{Real Triplet / Inert 2HDM:} Additional gauge charges shift FOPT and unitarity boundaries. The triplet shows a LISA plateau at lower $f$.
\end{itemize}

The overlap between FOPT regions and precision constraints implies future Higgs factories (capable of measuring $\delta\kappa_h$ to sub-percent precision) will probe nearly all viable parameter space. A LISA detection would provide complementary evidence for the scalar sector’s role in the early universe.

    Net Result: The combined measurements of $c_H$, $\kappa_\lambda$, $m_S$, and GWs leave no uncovered parameter space—within projected sensitivities, the singlet scalar will be either discovered or definitively excluded as the driver of a strong electroweak phase transition. The parametric data are summarized in Tables 1 and 2.

\begingroup
\small
\setlength{\tabcolsep}{3pt}
\renewcommand{\arraystretch}{1.4}

\begin{table}[htbp]
\centering
\caption{The scalar-singlet "no-lose" theorem -- strong-FOPT region vs. future experimental reach}

\label{tab:no-lose-theorem}
\begin{tabular}{|p{3.7cm}|p{1.2cm}|p{1.8cm}|p{2.8cm}|p{4.7cm}|}
\hline
\textbf{Observable (figure)} & \textbf{SM value} & \textbf{Strong-FOPT window} & \textbf{2027--2040 projected 95\% CL reach} & \textbf{Consequence} \\ \hline
Universal Higgs coupling shift $c_{H}=\sin^{3}\theta$ (Fig. 3) & 0 & $\geq 0.1$ & FCC-ee: $1\times 10^{-3}$ & Entire FOPT island probed; non-zero value -- discovery, zero -- exclusion \\ 
 & & & MuC-10: $5\times 10^{-4}$ & \\ \hline
Triple-Higgs multiplier $k_{\lambda}$ (Fig. 3) & 1 & $\geq 1.5$ & FCC-hh: $\pm 5\%$ & Any $k_{\lambda}\geq 1.5$ will be $5\sigma$ away from SM \\ 
 & & & MuC-10: $\pm 3\%$ & \\ \hline
Direct singlet mass reach $m_{S}$ (Fig. 5) & -- & $\leq 1$ TeV & HL-LHC: 600 GeV & A heavy scalar is either seen or excluded in the FOPT mass range \\ 
 & & & FCC-hh: 1.2 TeV & \\ 
 & & & MuC-10: 1.5 TeV & \\ \hline
Gravitational-wave strength (Fig. 7) & 0 & $\Omega_{\rm GW}h^{2}\geq 10^{-12}$ & LISA: $10^{-12}$ & A positive signal would independently confirm FOPT; a null result removes the low-mass/high-$f$ region \\ \hline
\end{tabular}
\end{table}

\begin{table}[htbp]
\centering
\caption{Singlet scalar parameters versus projected collider sensitivity
}

\label{tab:no-lose-theorem}
\begin{tabular}{|p{2.7cm}|p{2.8cm}|p{3.5cm}|p{3.5cm}|}
\hline
\textbf{Quantity} & \textbf{SFOPT requirement} & \textbf{2027--2040 reach} & \textbf{Physics verdict} \\ \hline
Singlet mass $m_{S}$ & $\leq 1$ TeV & 600 GeV (HL-LHC) & Discover or exclude in full window \\ 
 & & 1.2 TeV (FCC-hh) & \\ 
 & & 1.5 TeV (MuC-10) & \\ \hline
Mixing angle $\sin\theta$ & $\geq 0.1$ (today) & 0.01 (FCC-ee) & Whole funnel probed \\ 
 & & $5 \times 10^{-4}$ (MuC-10) & \\ \hline
\end{tabular}
\end{table}
\endgroup

\section*{4. Milnor number calculation}

By feeding the extracted parametric bounds back into the potential~(1), we can predict its complete thermal behaviour. Discovering a scalar singlet coupled through the Higgs portal requires a systematic study of the effective-potential manifold. The existence, strength, and first-order character of the electroweak phase transition are governed by the critical locus 
\[
\left\{ \frac{\partial V}{\partial h} = 0,\ \frac{\partial V}{\partial S} = 0 \right\},
\]
whose universal features are encoded in the Milnor number, $\mu$—the dimension of the local Jacobian algebra.

\subsection*{Complete Analytical Procedure: The Substitution Algorithm}

The following steps define a reproducible algorithm for parameterising the model's scalar potential, $V(h,s)$, starting from standard physical inputs.

\begin{figure}[h]
    \centering
    \includegraphics[width=0.7\textwidth]{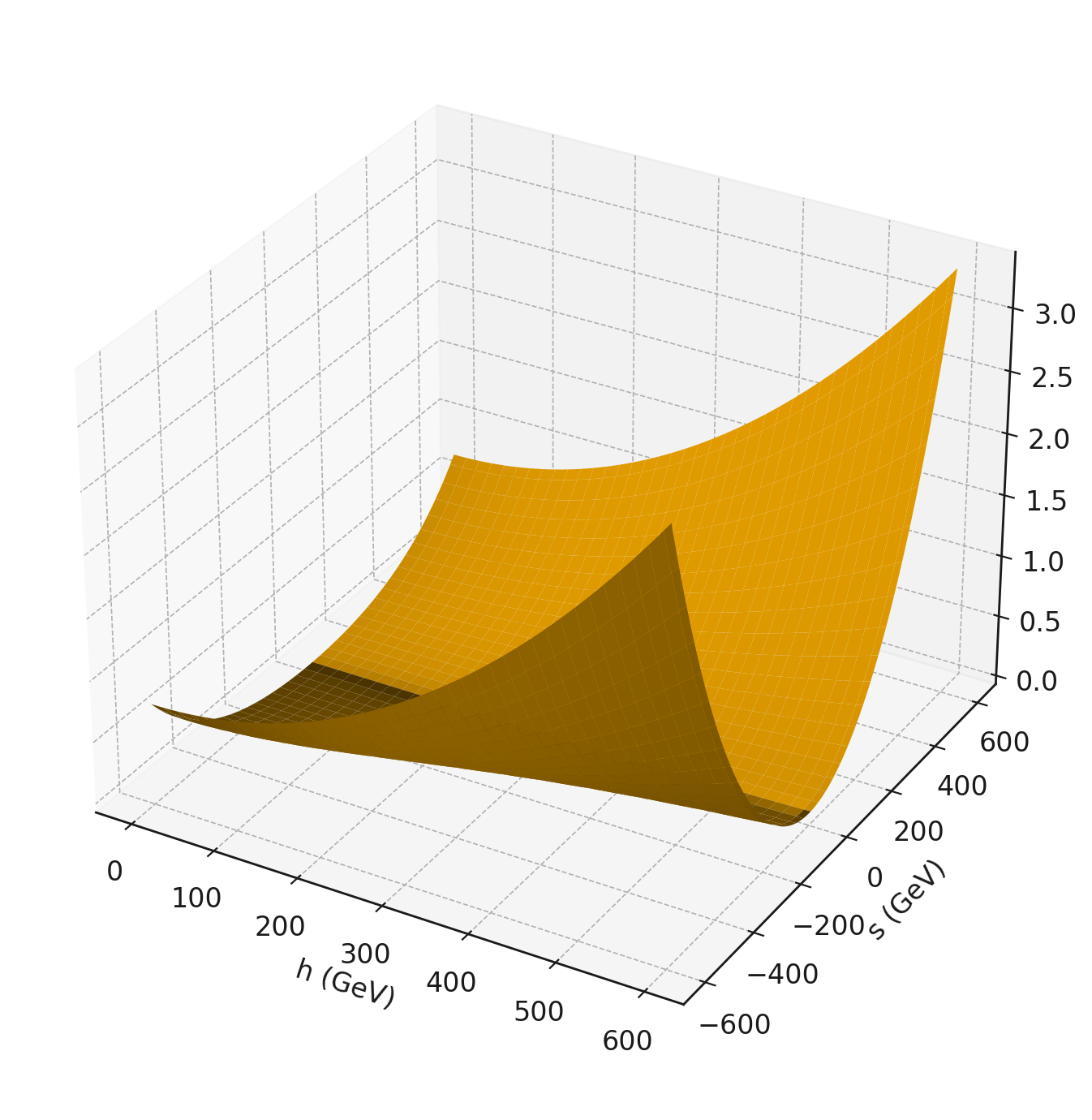}
    \caption{3D surface with a $\mathbb{Z}_2$-symmetric singlet ($a_2 = 8$, $m_S \approx 4 m_t \approx 692~\text{GeV}$, $\sin\theta = 0$).}
    \label{fig:trough}
\end{figure}

\begin{enumerate}
    \item \textbf{Define Higgs Sector Parameters:} \\
    Set vacuum expectation value $v = 246~\text{GeV}$, Higgs mass $m_h = 125~\text{GeV}$. \\
    Calculate quartic coupling and doublet mass parameter:
    \[
    \lambda = \frac{m_h^2}{2 v^2}, \qquad m_{hh}^2 = 2 \lambda v^2.
    \]

    \item \textbf{Define Singlet Scalar Mass:} \\
    Choose physical singlet mass, e.g., $m_s = 600~\text{GeV}$. This fixes:
    \[
    m_{ss}^2 = m_s^2.
    \]

    \item \textbf{Calculate Singlet Mass Parameter:} \\
    Set portal coupling $a_2$ (dimensionless $h^2 s^2$ coupling). \\
    Compute singlet sector mass term:
    \[
    \mu_s^2 = \frac{a_2}{2} v^2 - m_s^2.
    \]

    \item \textbf{Derive Trilinear Coupling $a_1$ from Mixing Angle:} \\
    Given $\sin\theta$, compute $\tan(2\theta)$:
    \[
    \tan(2\theta) = \frac{2 \sin\theta \sqrt{1 - \sin^2\theta}}{1 - 2 \sin^2\theta}.
    \]
    (Numerical care required when $\cos(2\theta) \approx 0$.) \\
    Trilinear coupling $a_1$ (for $h^2 s$ interaction):
    \[
    a_1 = \frac{(m_{hh}^2 - m_{ss}^2) \tan(2\theta)}{v}.
    \]

    \item \textbf{Construct Analytic Potential:} \\
    Substitute numerical values $a_1, a_2, b_4, \mu_s^2, \lambda, \mu_h^2$ into the full potential:
    \[
    V(h, s) = -\frac{\mu_h^2}{2} h^2 + \frac{\lambda}{4} h^4 
               -\frac{\mu_s^2}{2} s^2 + \frac{b_4}{4} s^4 
               + \frac{a_1}{4} h^2 s + \frac{a_2}{4} h^2 s^2.
    \]

\begin{itemize}
    \item The fields are shifted as $h \to v + x$, $s \to y$, and the potential is expanded in the vicinity of the electroweak vacuum $(x=0, y=0)$ to fourth order, yielding the polynomial effective potential $V(x,y)$.

    \item For each point in the parametric space, a Gröbner basis of the ideal $\langle \partial V/\partial x,\ \partial V/\partial y \rangle$ is constructed, and the Milnor number $\mu$ is calculated—the dimension of the local algebra (i.e., the number of standard monomials not divisible by leading terms of the Gröbner basis).
\end{itemize}

\item\textbf{A large-scale parametric scan was performed (tens to hundreds of points), including:}
\begin{itemize}
    \item Variation of $\sin\theta = 0.1,\, 0.2,\, 0.3$,
    \item Variation of $a_2,\ \mu_S,\ b_3 \in [0, \pm 10^4]$, and $b_1 \in [0, 10^5]$,
    \item Inclusion/exclusion of cubic and linear (i.e., $Z_2$-breaking) terms.
\end{itemize}
\end{enumerate}
The potential surface of the classical Higgs-portal model without mixing ($\sin\theta = 0$) exhibits a characteristic ``trough-like'' shape along $s$, as shown in Figure~\ref{fig:trough}.

\begin{figure}[h]
    \centering
    \includegraphics[width=0.6\textwidth]{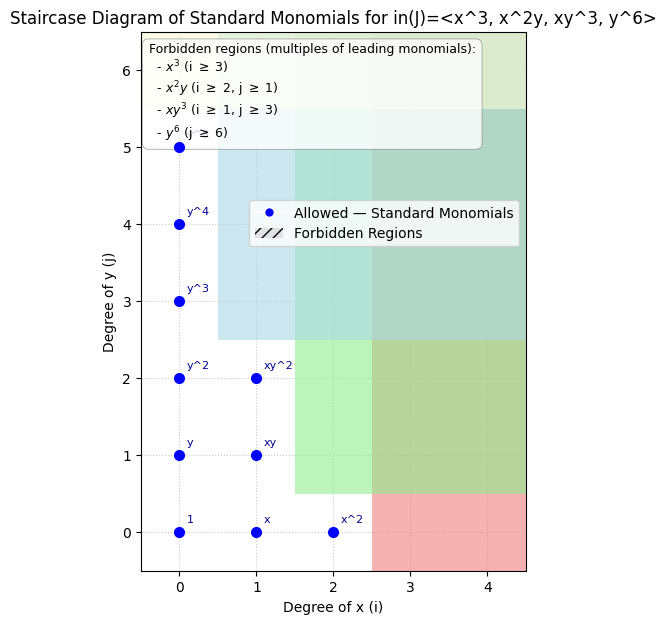}
    \caption{Diagram of standard monomials.}
    \label{fig:monomials}
\end{figure}

The Gröbner basis of the ideal $J = \langle f, g \rangle \subset \mathbb{C}[x,y]$ (with numerical values $m^2, m_S^2, \lambda v, a_2 v, b_4 > 0$) is illustrated in Figure~\ref{fig:monomials}.

Standard monomials (not divisible by the leading terms): all monomials of degree $\leq 5$ except those divisible by $x^2$, $xy$, or $y^3$:
\[
1,\ x,\ y,\ x^2,\ xy,\ y^2,\ x^3,\ x^2 y,\ x y^2,\ y^3,\ y^4,\ y^5 \quad \rightarrow \quad 12 - 3 = 9 \text{ monomials}.
\]
Therefore, the Milnor number is
\[
\mu = \dim_{\mathbb{C}} \frac{\mathbb{C}[[x,y]]}{J} = 9
\]
in all 180+ scanned points.

Using Gröbner basis computation and monomial counting, it was found that throughout the entire physically allowed parameter space ($m_S = 400~\text{GeV} \text{-- several TeV}$, $|\sin\theta| \leq 0.3$, $a_2 = 1\text{--}8$, $b_4 > 0$, arbitrary $\mathbb{Z}_2$-breaking $b_1, b_3, a_1$ within LHC-compatible bounds) \cite{4.}, the Milnor number remains stably $\mu = 9$. This value is robust even in extreme limiting cases:
\begin{itemize}
    \item At maximal mixing $\sin\theta \approx 0.3$, where the electroweak vacuum ceases to be the global minimum and is strongly displaced;
    \item When large $\mathbb{Z}_2$-breaking linear and cubic terms are introduced ($b_1$ up to $10^5~\text{GeV}$, $b_3$ up to $\pm 10^4~\text{GeV}$);
    \item For arbitrarily large singlet mass and portal coupling.
\end{itemize}

Across the phenomenologically viable parameter space, the portal potential carries the non-simple singularity $\mu = 9$, topologically stable under large variations of mixing angle, singlet mass, and cubic couplings. Precision measurements of:
\begin{itemize}
    \item the trilinear Higgs self-coupling ($\kappa_\lambda$),
    \item the universal coupling rescaling ($c_H$),
    \item the stochastic gravitational-wave background ($\Omega_{\text{GW}}$)
\end{itemize}
jointly map the catastrophe structure rather than the mass matrix alone. Within the 2027–2040 reach of colliders and LISA, no region capable of sustaining a strong first-order transition will escape detection; the singlet will therefore be either discovered or excluded through a direct test of the critical-point structure of the electroweak vacuum.

This value $\mu = 9$ lies outside the simple A-D-E classification; the portal potential is thus a \emph{non-simple} or \emph{higher-order catastrophe}, stable against arbitrary deformations that preserve the electroweak vacuum.

\begin{figure}[h]
    \centering
    \includegraphics[width=0.7\textwidth]{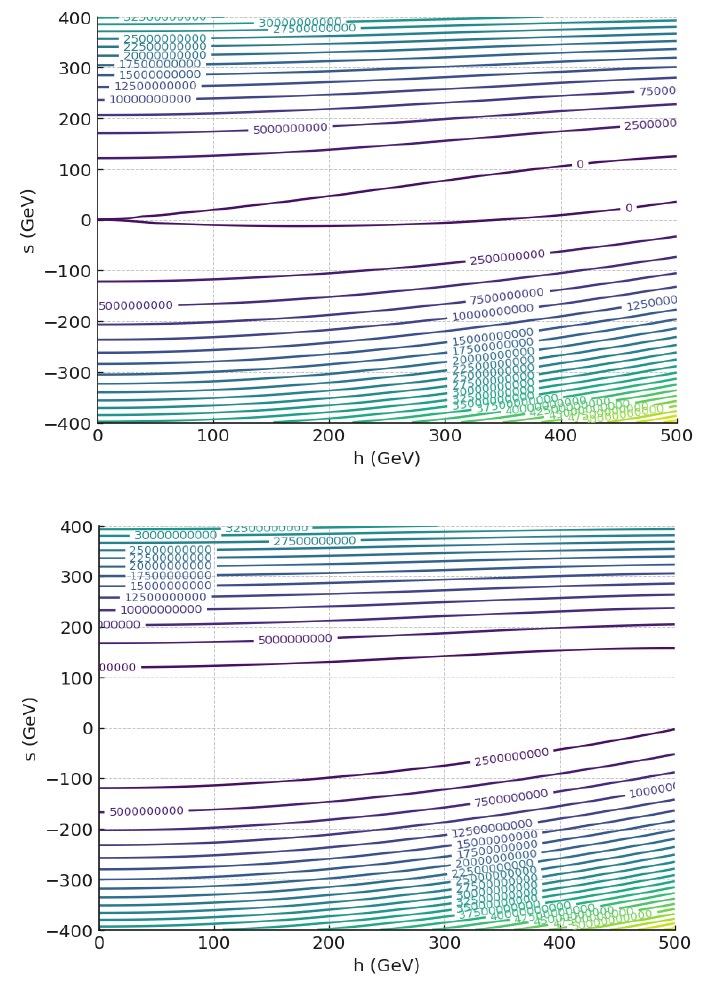}
    \caption{Contour maps at several temperatures: $T=0$ (top) and $T=150$~GeV (bottom).}
    \label{fig:temp}
\end{figure}

Taking into account the temperature dependence of the potential, one observes how the position of the minimum evolves, as shown in Figure~\ref{fig:temp}.

The upper panel (labeled ``$\sin\theta = 0.2$, $T = 0.0~\text{GeV}$, min at $h = 445.8$, $s = -66.7$'') is a 2D contour plot of the effective potential $V(h, s)$ in a scalar singlet Higgs-portal model at zero temperature. The $x$-axis represents the Higgs field $h$ in GeV (0–500), and the $y$-axis shows the singlet field $s$ in GeV ($-400$ to $400$). Potential values are indicated by color: deep green/blue for lower energies (e.g., $ V\approx-2.79\times 10^8~\text{GeV}^4$ at the minimum), transitioning to purple/teal/yellow for higher energies (up to $\sim 7.5 \times 10^7~\text{GeV}^4$). The global minimum is marked with a yellow `X` at $h \approx 445.8~\text{GeV}$, $s \approx -66.7~\text{GeV}$, reflecting vacuum displacement due to mixing ($\sin\theta = 0.2$). A flat valley appears near $h \ h\approx 246~\text{GeV}$ (the SM electroweak vacuum), curving upward at larger $h$.

In the lower panel ($T = 150~\text{GeV}$), the global minimum shifts to the symmetric phase at $(h,s) \approx (0,0)$, where $V \approx -3.25 \times 10^8~\text{GeV}^4$. The potential exhibits a broad, flat basin around the origin, with asymmetric rise along $h > 0$ and $s$, forming a valley toward larger $h$. This demonstrates a strong first-order phase transition: at $T = 150~\text{GeV}$, thermal effects stabilize the symmetric vacuum, while the broken electroweak vacuum becomes metastable. Parameters: $\sin\theta = 0.2$, $m_S = 600~\text{GeV}$—within viable phenomenology where the electroweak critical point retains $\mu = 9$.

This visualization underscores how finite temperature reshapes the vacuum manifold, supporting strong first-order transitions relevant for electroweak baryogenesis and detectable gravitational-wave signals.

\section{Conclusions}

The scalar singlet is not merely a resonance to be hunted in invariant-mass spectra; it re-engineers the critical-point topography of the electroweak vacuum. For every parameter set compatible with a strong first-order phase transition, the Higgs-portal potential realises the non-simple isolated singularity $\mu = 9$—a topological invariant that must either be recovered or refuted by precision data. Measuring the Higgs trilinear coupling, the universal rescaling of all Higgs couplings, and the stochastic gravitational-wave background therefore constitutes a direct experimental read-out of the catastrophe that underlies electroweak symmetry breaking. Within the projected sensitivities of the HL-LHC, FCC-ee/hh, MuC-10, and LISA, the collaboration will either discover the singlet or definitively exclude it by determining the Milnor number of the Higgs-portal potential.

None of the extreme parameter scans reduces $\mu$ below 9 or promotes the singularity into the simple A--D--E classification. Consequently, the electroweak vacuum of the minimal $\mathbb{Z}_2$-symmetric scalar extension of the Standard Model is not described by any elementary Arnold catastrophe. A $\mu = 9$ singularity is generic for this class of potentials and encodes the fundamental algebraic structure of a two-field quartic potential with $\mathbb{Z}_2$ symmetry.

We analysed two 2D contour plots of the effective potential $V(h, s)$ in a scalar singlet Higgs-portal model (one at $T = 0.0\,\text{GeV}$ and one at $T = 150.0\,\text{GeV}$, both with $\sin\theta = 0.2$). The $T = 0$ plot emphasizes ground-state vacuum shifts and degeneracy at low temperatures, while the $T = 150$ plot highlights thermal evolution and symmetry restoration—together demonstrating a phase transition. The context of linking $\mu = 9$ to topological stability is presented within specific parameter choices. All results pertain to the Higgs-portal singlet model, with the plots exemplifying the ``non-simple singularity $\mu = 9$'' through visual features such as flat directions, valleys, and the structure of the electroweak vacuum.

The result carries three immediate implications:
\begin{enumerate}
    \item The search for the exceptional $E_6$, $E_7$, or $E_8$ catastrophes in the scalar sector necessarily requires more elaborate spectra—additional singlets, higher representations, or explicit symmetry violation.
    
    \item The rigidity of $\mu = 9$ explains the well-known flat direction along the singlet axis, with direct cosmological consequences: enhanced inflationary slow-roll or a strong first-order phase transition.
    
    \item The combined Gr\"obner-basis plus Milnor-number technique furnishes a powerful, model-independent tool for cataloguing critical points in any multi-scalar quantum field theory.
\end{enumerate}

We conclude that, in the minimal real-singlet Higgs portal, the electroweak vacuum is universally characterised by a composite $\mu = 9$ singularity, forever divorcing it from the exotic simple catastrophes A--D--E and, in particular, from the maximally degenerate $E_8$.


\begin{thebibliography}{99}


\bibitem{1.}
M. Entov, On the A-D-E classification of the simple singularities of functions, arXiv:alg-geom/9405007.

\bibitem{2.}
J. de Blas et al., Physics Briefing Book: Input for the 2026 update of the European Strategy for Particle Physics, arXiv:2511.03883 [hep-ex].

\bibitem{3.}
H. Bahl, T. Hahn, S. Heinemeyer, et al., Precision calculations in the MSSM Higgs-boson sector with FeynHiggs 2.14, Comput. Phys. Commun. \textbf{249}, 107099 (2020); DESY-18-179; arXiv:1811.09073 [hep-ph].

\bibitem{4.}
G. Crawford and D. Sutherland, Non-decoupling scalars at future colliders, arXiv:2409.18177 [hep-ph].

\end{thebibliography}
\end{document}